\documentclass[prb,aps,twocolumn,superscriptaddress,showpacs]{revtex4}

\usepackage{amsfonts}
\usepackage{graphicx}
\usepackage{amsmath}

\begin{document}

\title{ Spontaneous symmetry breaking in a generalized orbital compass model }

\author{     Lukasz Cincio}
\affiliation{Marian Smoluchowski Institute of Physics
             and Centre for Complex Systems Research,  \\
             Jagiellonian University, Reymonta 4, 30-059 Krak\'ow, Poland}

\author{     Jacek Dziarmaga}
\affiliation{Marian Smoluchowski Institute of Physics
             and Centre for Complex Systems Research,  \\
             Jagiellonian University, Reymonta 4, 30-059 Krak\'ow, Poland}

\author{     Andrzej M. Ole\'s}
\affiliation{Marian Smoluchowski Institute of Physics
             and Centre for Complex Systems Research,  \\
             Jagiellonian University, Reymonta 4, 30-059 Krak\'ow, Poland}
\affiliation{Max-Planck-Institut f\"{u}r Festk\"{o}rperforschung,
             Heisenbergstrasse 1, 70569 Stuttgart, Germany}

\date{ 4 July 2010 }

\begin{abstract}
We introduce a generalized two-dimensional orbital compass model, which 
interpolates continuously from the classical Ising model to the orbital 
compass model with frustrated quantum interactions, and investigate
it using the multiscale entanglement renormalization ansatz (MERA).
The results demonstrate that increasing frustration of exchange
interactions triggers a second order quantum phase transition to a
degenerate symmetry broken state which minimizes one of the
interactions in the orbital compass model. Using boson expansion
within the spin-wave theory we unravel the physical mechanism of
the symmetry breaking transition as promoted by weak quantum
fluctuations and explain why this transition occurs only surprisingly
close to the maximally frustrated interactions of the orbital compass 
model. The spin waves remain gapful at the critical point, and both 
the boson expansion and MERA do not find any algebraically decaying 
spin-spin correlations in the critical ground state.\\
{\it Published in: Physical Review B {\bf 82}, 104416 (2010).}
\end{abstract}


\maketitle

\section{Introduction}

The orbital compass model (OCM) is physically motivated by the orbital
interactions which arise for strongly correlated electrons in
transition metal oxides with partly filled degenerate $3d$ orbitals
and lead to rich and still poorly understood quantum models. In
these systems the orbital degrees of freedom play a crucial role in
determining collective states such as coexisting magnetic and
orbital order, as for instance in the colossal magnetoresistance
manganites\cite{Tok00} or in the vanadate perovskites.\cite{Hor08}
The orbital interactions are typically intrinsically frustrated and
may strongly enhance quantum fluctuations, leading to disordered
states. \cite{Fei97} While realistic orbital interactions are
somewhat complex, a paradigm of intrinsic frustration is best
realized in the OCM,\cite{Kho03,Nus04,Mis04,Dor05,Dou05} with the
pseudospin couplings intertwined with the orientation of interacting
bonds. Its two-dimensional (2D) version on a honeycomb lattice,
\cite{Jac09} realized in layered iron oxides,\cite{Nag07} is equivalent
to the Kitaev model.\cite{Kit06}

Although conceptually quite simple, the OCM has an interdisciplinary
character as it plays an important role in a variety of contexts
beyond the correlated transition metal oxides, such as: (i) the
implementation of protected qubits for quantum computations in
Josephson lattice arrays,\cite{Dou05} (ii) topological quantum
order,\cite{Nus09} or (iii) polar molecules in optical lattices and
systems of trapped ions.\cite{Mil07} Numerical studies \cite{Dor05}
suggested that when anisotropic interactions are varied through the
isotropic point of the 2D OCM, the ground state is not an orbital
liquid type but instead a first order quantum phase transition (QPT)
occurs between two different types of Ising-type order dictated by
one or the other interaction. Recently the existence of this
transition, similar to the one which occurs in the exact solution of
the one-dimensional OCM,\cite{Brz07} was confirmed using projected
entangled-pair state algorithm.\cite{Oru09} This implies that the
symmetry is spontaneously broken at the compass point, and the spin
order follows one of the two equivalent frustrated interactions.

Knowing that the ground states of the 2D Ising model and the 2D OCM are
quite different, we introduce a generalized OCM which interpolates
between these two limiting cases. Using this model we will
investigate: (i) the physical consequences of gradually increasing
frustration in a 2D system, (ii) where a QPT occurs from the Ising
ground state to the degenerate ground state of the OCM, and,
finally, (iii) the order and the physical mechanism of this QPT. As
increasing frustration of the orbital interactions introduces
entangled states, the present problem provides a unique opportunity
to use the recently developed multiscale entanglement renormalization 
ansatz\cite{Vid07,Vid08} (MERA) in order to find reliable answers to 
the above questions. As we show below, the QPT in the generalized OCM 
occurs only surprisingly close to the maximally frustrated interactions 
in the OCM. We also explain the physical origin of this behavior using 
an analytic approach based on the spin-wave theory.

Quantum many-body systems exhibit several interesting collective
phenomena. Recent progress in developing efficient numerical methods
to study quantum systems on a lattice is remarkable and allowed to
investigate complex many-body phenomena, including
QPTs.\cite{Sac99} An important step here was the
discovery of density matrix renormalization group, \cite{Whi92} a
very powerful numerical method that can be applied to
one-dimensional strongly correlated fermionic and bosonic
systems.\cite{Hal06} This idea played a fundamental role in
developing entanglement renormalization \cite{Vid07,Vid08} to study
quantum spin systems on a 2D lattice. Crucial in
this approach is the removal of short-range entanglement by unitary
transformations called disentanglers. It generates a real-space
renormalization group transformation implemented in the MERA, which
was recently successfully employed to investigate several quantum
spin models,\cite{Cin08,EveIsing,Eve09,Gio,Dav08} and interacting
fermion systems.\cite{Eis09,Vid10} So far, the very
promising MERA has been applied {\it inter alia\/} to the 2D quantum
Ising model,\cite{Cin08,EveIsing} and to the Heisenberg model
on a kagome lattice,\cite{Kagom} but other possible applications
and the optimal geometries for performing sequentially
disentanglement and isometry transformation were also
discussed.\cite{Eve09}

The paper is organized as follows. In Sec. II we introduce the
generalized OCM and state the problem of the existence and nature of the
QPT. Next we present the MERA algorithm in Sec. \ref{sec:mera} used
to investigate frustrated interactions in the OCM. Numerical results
obtained using the MERA are presented in Sec. \ref{sec:num}. In
order to explain the physical mechanism of the QPT found in the OCM
we performed the boson expansion within the spin-wave theory, as
described in Sec. \ref{sec:lsw}. The details of this expansion are
presented in the Appendix. The paper is summarized in Sec.
\ref{sec:summa}, where the main conclusions of the present work are
also given.

\section{Generalized compass model}
\label{sec:geco}

\begin{figure}[t!]
\includegraphics[width=8.4cm,clip=true]{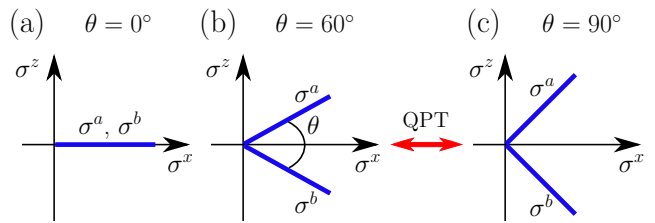}
\caption{(Color online) Artist's view of the evolution of orbital
interactions in the generalized OCM Eq. (\ref{H}) with increasing
angle $\theta$. Heavy (blue) lines indicate favored spin direction induced
by interactions along two nonequivalent lattice axes $a$ and $b$.
Different panels show: (a) the Ising model at $\theta=0^\circ$, (b)
the 2D $e_g$ orbital model at $\theta=60^\circ$, and (c) the OCM at
$\theta=90^\circ$. Spin order follows the interactions in the Ising
limit, while it follows one of the equivalent interactions, 
$\sigma^a$ or $\sigma^b$, in the OCM. This results in the symmetry 
breaking QPT which occurs between (b) and (c), as we show in Secs. 
\ref{sec:num}-\ref{sec:lsw}.}
\label{fig:ocm}
\end{figure}

In this paper, we investigate the nature and position of the QPT
when the OCM point is approached in a different way from that
studied before,\cite{Dor05,Oru09} namely when frustration of
interactions along two nonequivalent directions gradually increases.
Therefore, we introduce a 2D {\it generalized\/} OCM with ferro-like
interactions \cite{notefo} on a square lattice in $ab$ plane (we
assume the exchange constant $J=1$),
\begin{equation}
{\cal H}(\theta)=-\!\sum_{ij\in ab} \!\Big\{
\sigma^a_{ij}(\theta)\sigma^a_{i+1,j}(\theta) +
\sigma^b_{ij}(\theta)\sigma^b_{i,j+1}(\theta) \Big\}\,. \label{H}
\end{equation}
The interactions occur between nearest neighbors and are balanced
along both lattice directions $a$ and $b$. Here $\{ij\}$ labels
lattice sites, with $i$ ($j$) increasing along $a$ ($b$) axes, and
$\{\sigma^a_{ij}(\theta),\sigma^b_{ij}(\theta)\}$ are linear
combinations of Pauli matrices describing interactions for $S=1/2$
spins:
\begin{eqnarray}
\sigma^a_{ij}(\theta)\! &=& \cos(\theta/2)\;\sigma^x_{ij}
                           +\sin(\theta/2)\;\sigma^z_{ij}\,,\\
\sigma^b_{ij}(\theta)\! &=& \cos(\theta/2)\;\sigma^x_{ij}
                           -\sin(\theta/2)\;\sigma^z_{ij}\,.
\label{sigmaab}
\end{eqnarray}
The interactions in Eq. (\ref{H}) include the classical Ising model
at $\theta=0^\circ$ for $\sigma^x_{ij}$ operators and become
gradually more frustrated with increasing angle
$\theta\in(0^\circ,90^\circ]$ --- they interpolate between the Ising
model (at $\theta=0^\circ$) and the isotropic OCM (at
$\theta=90^\circ$), see Fig. \ref{fig:ocm}. The latter case is
equivalent to the 2D OCM with standard interactions
$\sigma^z_{ij}\sigma^z_{i,j+1}$ and $\sigma^x_{ij}\sigma^x_{i+1,j}$
along the $a$ and $b$ directions
\cite{Kho03,Nus04,Mis04,Dou05,Dor05} by a straightforward unitary
transformation. The model (\ref{H}) includes also as a special case
the 2D orbital model for $e_g$ electrons at
$\theta=60^\circ$,\cite{notefo} describing, for instance, the
orbital part of the superexchange interactions in the ferromagnetic
planes of LaMnO$_3$.\cite{Fei99}

Since the {\it isotropic} model has the same interaction strength for the bonds
along both $a$ and $b$ axis, it is symmetric under transformation
$a\leftrightarrow b$, and the issue of the QPT between different ground
states of the {\it anisotropic} compass model\cite{Oru09} does not arise.
On one hand, this symmetry is obeyed by
the classical Ising ground state, while on the other hand, in the
ground state of the OCM this symmetry is spontaneously broken (and
the ground state is degenerate).
Therefore, an intriguing question concerning the ground state of the
model (\ref{H}) is whether it has the same high symmetry as the
Ising model in a broad range of $\theta$, or the symmetry is soon
spontaneously broken when $\theta$ increases, i.e., there are
degenerate ground states with lower symmetries, also for the $e_g$
orbital model, see Fig. \ref{fig:ocm}(b). This question has been
addressed by investigating the energy contributions along two
equivalent lattice directions $a$ and $b$ by applying the MERA.

\section{MERA Algorithm}
\label{sec:mera}

\subsection{Calculation method}

\begin{figure}[t!]
\includegraphics[width=7cm,clip=true]{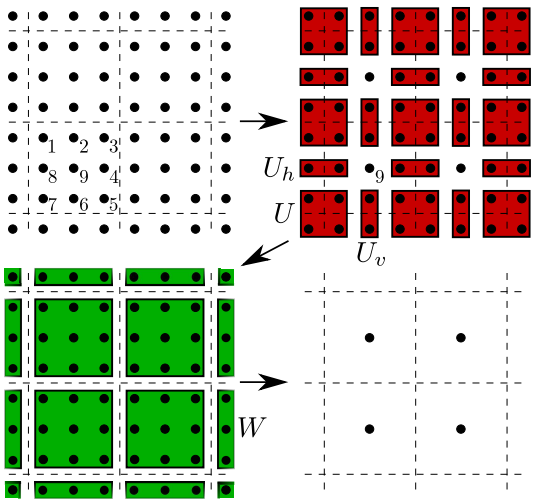}
\caption{(Color online) 9-to-1 geometry of MERA applied to the OCM:
dark (red) boxes  represent the action of disentanglers $U, U_h,
U_v$ and gray (green) ones --- isometries $W$; arrows indicate
subsequent transformations used; the labels of spins 1-9 in a single
block are addressed in the text. This is a coarse-graining
renormalization group transformation where each $3\times 3$
plaquette in the top-left panel is replaced by a coarse grained spin
in the bottom-right panel (nine spins are replaced by
one coarse-grained spin). To minimize the number of states $\chi$ of the
coarse-grained spin, the microscopic spins are disentangled prior to
decimation. We also used a 5-to-1 geometry,
see Fig. 1 of Ref. \onlinecite{EveIsing}.
}
\label{fig:geo}
\end{figure}

In order to obtain the ground state, we use a translationally
invariant MERA on infinite lattice.\cite{Eve09} The MERA is a tensor
network with infinite number of layers of disentanglers and
isometries. By translational invariance, all isometries
(disentanglers) in a given layer are the same. Since every layer
represents a coarse-graining renormalization group transformation,
shown in Fig. \ref{fig:geo} for the 9-to-1 geometry and described in
more detail in Ref. \onlinecite{Eve09} (see their Fig. 7), we assume that
after a finite number of such transformations a fixed point of the
renormalization group is reached (either trivial or non-trivial) and
from that time on the following transformations are the same. In
other words, at the bottom of the tensor network there is a finite
number of non-universal layers whose tensors are different in
general, but above certain level all layers are the same. The bottom
layers describe non-universal short range correlations, and the
universal layers above this level describe universal properties of
the fixed point. The number $N$ of the non-universal bottom layers
is one of the parameters of the infinite-lattice MERA. We have
verified that it is enough to keep up to three non-universal layers,
depending on how close the critical point is.

Starting with randomly chosen tensors, the structure is optimized
layer by layer, from the top to bottom and back. In given layer
$\tau$, we calculate an environment of each tensor type by means of
renormalized Hamiltonians $h_\tau$ and density matrices $\rho_\tau$
computed from other layers. The environments are aimed at updating
tensors to minimize total energy. In the universal layer, this
updating technique is slightly different: $h_\infty$ and
$\rho_\infty$ are fixed points of the renormalization procedure
defined by tensors in this layer. The above steps are iterated until
the convergence of energy is achieved. For given $\theta$, we obtain
the ground state for different values of bond dimension $\chi$. It
turns out that in most cases it is sufficient to work with $\chi=3$,
which is the same in each layer. However, it is necessary to
increase $\chi$ to $4$ in the neighborhood of the critical
point. The number of operations and the required memory scale as
$\chi^{16}$ and $\chi^{12}$ respectively. The inset in Fig.
\ref{fig:meraMF}(a) shows the convergence of the energy of the ground
state with an increasing bond dimension. Here we also present a
comparison of results obtained with the alternative 5-to-1 geometry.

The algorithm is implemented in \verb+c+++ and optimized in order to
work on multi-processor computers. On an eight-core 2.3 GHz
processor, it takes about half an hour to update the whole tensor
network which consists of four layers of tensors with $\chi=4$. Near
the critical point, i.e., at $\theta\simeq\theta_c$, the convergence
requires several thousands of iterations whereas it is significantly
faster far from $\theta_c$. When $\theta$ is scanned from $0^\circ$
to $90^\circ$ (or back), it is more efficient to use the previous
ground state as an initial state for the next discrete value of
$\theta$ instead of starting from a random initial state for each
value of $\theta$. We have carefully verified convergence to the
ground state by scanning $\theta$ back and forth and comparing the
results with those obtained from random initial states for selected
values of $\theta$.

\subsection{Correlations}

\begin{figure}[t!]
\includegraphics[width=8.2cm]{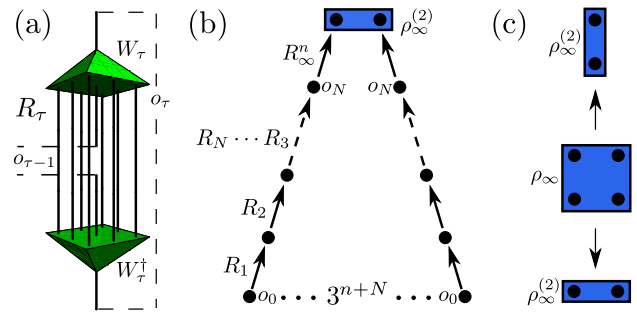}
\caption{(Color online) (a) Renormalizer superoperator which
consists of isometries only. The connections show how the isometries
are contracted; compare Eq. (\ref{R}). (b) Method of calculating
correlations $\langle o_{\bf x} o_{\bf y} \rangle = \langle o_0 o_0
\rangle$ between two sites separated by a distance $3^{n+N}$. The
scheme presents a graphical explanation of Eq. (\ref{Tr}). (c)
Deriving $\rho_\infty^{(2)}$ from $\rho_\infty$ when sites are
separated vertically (top) and horizontally (bottom). }
\label{fig:superop}
\end{figure}

In order to calculate correlations, we take advantage of the special
structure of the renormalization group transformation in Fig.
\ref{fig:geo}. A site of the lattice that lies in the center of a $3
\times 3$ decimation block (number 9 in Fig. \ref{fig:geo})
undergoes renormalization in a particularly easy manner. Since no
disentangler is applied to this central site, a one-site operator
$o_{\tau-1}$ at this site is mapped by the $\tau$-th renormalization
group transformation to a coarse-grained one-site operator
\begin{equation}
o_\tau ~=~ R_\tau o_{\tau-1} \, .
\end{equation}
Here $R_\tau$ is a renormalizer superoperator built out of
contracted isometries only, as shown in Fig. \ref{fig:superop}(a):
\begin{equation}
(R_\tau)_{kl}^{ij} ~=~ \sum_{n_1,\ldots, n_8} (W_\tau)^i_{n_1
\ldots n_8 k} \ ( W^\dag_\tau )_{j}^{n_1 \ldots n_8 l} \,.
\label{R}
\end{equation}
The meaning of the transformations $W_\tau$ and $W_\tau^{\dagger}$ is given in Figs.
\ref{fig:geo} and \ref{fig:superop}(a).
Thus, if we have $N$ non-universal layers at the bottom of the
geometry of MERA, then renormalized one-site operators at the
central sites just below the universal layer are given by [{see
Fig. \ref{fig:superop}(b)]:
\begin{equation}
o_N ~=~ R_N R_{N-1} \cdots R_1 o_0 \, , \label{oN}
\end{equation}
where $o_0 \equiv o$ denotes a physical, microscopic one-site
operator at one of the central sites at the very bottom of the
MERA tensor network.

To extract information on the correlations, it is convenient to
write eigen-decomposition of the renormalizer $R_\infty$ in the
universal layer:
\begin{equation}
R_\infty v_\alpha ~=~ \lambda_\alpha v_\alpha \, .
\end{equation}
It is straightforward to verify the basic property of the spectrum
of $R_\infty$: $|\lambda_\alpha| \leq 1$. The ortonormality of the
vectors $W^i$ in Eq. (\ref{R}) implies that the identity operator
$(v_1)_{ij}=\delta_{ij}$ is an eigenvector with eigenvalue
$\lambda_1 = 1$. In our numerical calculations this is the only
eigenvalue with modulus $1$.

After the operator $o_N$ is decomposed as $o_N = \sum_\alpha
o_N^\alpha v_\alpha$, a repeated action of the renormalizer
$R_\infty$ in the universal layers can be written as
\begin{equation}
R_\infty^n o_N ~=~ \sum_\alpha \lambda_\alpha^n \, o_N^\alpha \,
v_\alpha \, .
\end{equation}
A correlator between two central sites ${\bf x}$ and ${\bf y}$
separated by a distance $|{\bf x} - {\bf y}| = 3^{n+N}$ in the
horizontal (vertical) direction is thus given by:
\begin{eqnarray}
\langle o_{\bf x} o_{\bf y} \rangle &=& \textrm{Tr} \left\{
\rho_\infty^{(2)}\, (R_\infty^n o_N \otimes R_\infty^n o_N) \right\}
\label{Tr} \\
 &=& \sum_{\alpha,\beta} o_N^\alpha o_N^\beta \, c_{\alpha \beta}
 \lambda_\alpha^n \lambda_\beta^n \\
 &=& \sum_{\alpha,\beta} \frac{o_N^\alpha o_N^\beta \,
 c_{\alpha \beta}}{r^{-\log_3 (\lambda_\alpha \lambda_\beta)}} \, ,
\label{corr}
\end{eqnarray}
where $r = 3^n$ and
\begin{equation}
c_{\alpha\beta} = \textrm{Tr}\left\{ \rho^{(2)}_\infty ~ (v_\alpha
\otimes v_\beta)\right\}\,.
\end{equation}
Here $\rho^{(2)}_\infty$ is a two-site reduced density matrix in a
universal layer derived from $\rho_\infty$ as depicted in Fig.
\ref{fig:superop}(c).

Correlations corresponding to the leading eigenvalue $\lambda_1=1$
do not decay with the distance between ${\bf x}$ and ${\bf y}$. They
describe long range order in the operator $o$ and can be used to
extract its expectation value $\langle o\rangle$:
\begin{equation}
\langle o \rangle^2 ~=~ \lim_{|{\bf x} - {\bf y}| \to \infty}
\langle o_{\bf x} o_{\bf y} \rangle ~=~ o_N^1 o_N^1 c_{11} ~=~
\left(o_N^1\right)^2\, , \label{LRO}
\end{equation}
where we use the property: $\lim_{n \to \infty} \lambda^n_\alpha =
0$ that holds for $\alpha > 1$ and the fact that $c_{11} = 1$ which
is a consequence of $v_1$ being an identity. Thus only a one-site
operator with a non-zero coefficient $o_N^1$ has non-zero
expectation value. A trivial example is the identity $o =
\mathbb{I}$. Indeed, we obtain $o_N=\mathbb{I}$ in Eq. (\ref{oN}),
which is equivalent to $o_N^1=1$, and Eq. (\ref{LRO}) yields
$\langle  \mathbb{I} \rangle^2=1$ as expected.

\section{Numerical results}
\label{sec:num}

\subsection{Symmetry breaking transition}

Information about the ground state of the OCM Eq. (\ref{H}) is
contained in average energy per bond $E(\theta)$ and energy
anisotropy $\Delta E(\theta)$:
\begin{eqnarray}
\label{E} E(\theta)\!\!&=&\!-\frac12\Big\langle
\sigma^a_{ij}(\theta)\sigma^a_{i+1,j}(\theta)+
\sigma^b_{ij}(\theta)\sigma^b_{i,j+1}(\theta)\Big\rangle ,\\
\label{dE} \Delta E(\theta)\!\!&=&\!\Big|\left\langle
\sigma^a_{ij}(\theta)\sigma^a_{i+1,j}(\theta)\rangle-
\langle\sigma^b_{ij}(\theta)\sigma^b_{i,j+1}(\theta)\right\rangle\Big|.
\end{eqnarray}
In the classical limit of Ising interactions $E(0^\circ)=-1$ and $\Delta
E(0^\circ)=0$. Due to increasing frustration, the energy $E(\theta)$
gradually increases for increasing angle $\theta$ in Eq. (\ref{H})
and reaches a maximum of $E(90^\circ)\simeq -0.57$ in the OCM, see
Fig. \ref{fig:meraMF}(a). This increase is smooth and does not
indicate the existence of a QPT.

However, by investigating the anisotropy $\Delta E(\theta)$ Eq.
(\ref{dE}) between $a$ and $b$ bonds, we identified an angle
$\theta_c$ at which $\Delta E(\theta)$ starts to grow. Although a
gradual evolution of the ground state staring from $\theta=0^\circ$
might be also expected, the Ising-type state is first surprisingly
robust in a broad range of angles $\theta\in[0^\circ,\theta_c]$, and the
energy associated with bonds along the $a$ and $b$ axes remains the
same, i.e., $\Delta E(\theta)\equiv 0$. Next, the symmetry between
the $a$ and $b$ directions is spontaneously broken above $\theta_c$, 
where a finite value of $\Delta E(\theta)$ is found, and then $\Delta E(\theta)$ 
grows rapidly with further increasing angle $\theta$, i.e., large spin 
correlations develop along only one of the two equivalent directions $a$ 
and $b$. This QPT was detected by the MERA at $\theta_c\simeq 84.8^\circ$,
see Fig. \ref{fig:meraMF}(b).

\begin{figure}[t!]
\includegraphics[width=8.2cm,clip=true]{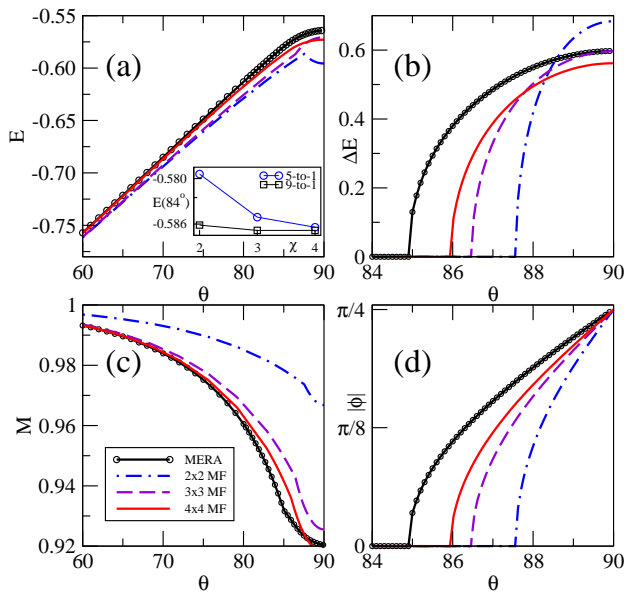}
\caption{(Color online) Ground state obtained for the generalized
OCM Eq. (1) using the MERA: (a) average energy $E$ per bond given by Eq.
(\ref{E}), (b) energy anisotropy $\Delta E$ given by Eq. (\ref{dE}), (c)
spontaneous magnetization $M$ given by Eq. (\ref{Mlocal}), and (d)
magnetization orientation $\phi$ given by Eq. (\ref{arctan}). Embedded
$L\times L$ clusters coupled to the neighboring spins  by mean-field
terms ($L\times L$ MF) exhibit qualitatively similar behavior.
\textit{Inset}: Convergence of the ground state energy obtained by two
geometries of MERA with increasing bond dimension $\chi$. Black:
9-to-1 geometry presented in Fig. \ref{fig:geo}; blue: 5-to-1
geometry introduced in Fig. 1 of Ref. \onlinecite{EveIsing}. The
9-to-1 geometry results prove to converge faster for $\theta$ close
to $\theta_c$.\cite{note} } \label{fig:meraMF}
\end{figure}

\subsection{Magnetization in the ground state}\label{SectionM}

To understand better the QPT at $\theta_c$ let us consider the
expectation value of the spontaneous magnetization 
${\bf M}\equiv\{M^x,M^y,M^z\}$ derived from the long range
order in the correlation function:
\begin{equation}
\lim_{|{\bf x-y}|\to\infty}
\langle\sigma^k_{\bf x}\sigma^l_{{\bf y}}\rangle = M^k M^l\,,
\label{M}
\end{equation}
where $k(l)=x,y,z$. For the interactions in Eq. (1) one finds $M^y\equiv 0$
for any $\theta$.

We found that the ground state obtained using the MERA for $\theta<\theta_c$ 
is characterized by $M^z=0$ and Ising-type long range order of $M^x$ which 
gradually decreases but remains rather large, $|M^x|>0.93$, in this parameter 
range. The symmetry between the directions $a$ and $b$ is broken above $\theta_c$ 
by appearance of a nonzero component $M^z$.

The value of the total magnetization
\begin{equation}
M=|{\bf M}|\equiv\sqrt{(M^x)^2+(M^z)^2}\,,
\label{Mlocal}
\end{equation}
obtained from the MERA decreases continuously from $M(0^\circ)=1$ in
the Ising model to $M(90^\circ)\simeq 0.92$ in the OCM, see Fig.
\ref{fig:meraMF}(c). Thus the reduction in the order parameter $M$ by
quantum fluctuations arising from the admixture of the $z$-th
component, is here rather small, and reproduces qualitative results
obtained for the $e_g$ orbital model within the linear orbital wave
theory \cite{vdB99}. Furthermore, by a closer inspection of
$M(\theta)$ we have found that the derivative $(\partial
M(\theta)/\partial\theta)$ does not exist at $\theta=\theta_c$.

As expected from the behavior of $\Delta E$, the obtained symmetry
breaking shown in Fig. \ref{fig:meraMF} implies that the direction
of spontaneous magnetization $\bf{M}$, parametrized by an
orientation angle
\begin{equation}
\phi~=~\arctan\left(\frac{M^z}{M^x}\right)\,, \label{arctan}
\end{equation}
begins to change when $\theta$ increases above $\theta_c$, see Fig.
\ref{fig:meraMF}(d). For $\theta<\theta_c$, the magnetization has
only one component $M^x\neq 0$ with $\phi=0$, pointing either
parallel or anti-parallel to $\sigma^x$ which is half-way between
$\sigma^a(\theta)$ and $\sigma^b(\theta)$, see Fig. \ref{fig:ocm}.
Below $\theta_c$ the ferromagnetic ground state is doubly degenerate
and the magnetization is $\pm M=\pm|M^x|$. When $\theta$ increases
above $\theta_c$ the magnetization begins to rotate in the
$\{M^x,M^z\}$ plane by the non-zero angle $\pm\phi$ Eq. (\ref{arctan})
with respect to the $\pm|M^x|$ initial magnetization
below $\theta_c$, and each of these two states splits off into two
ferromagnetic states rotated by $\pm|\phi|$ with respect to the
$\sigma^x$-axis. As a result, one finds four degenerate states above
$\theta_c$, and each of them is tilted with respect to
$\pm\sigma^x$, either toward $\pm\sigma^a(\theta)$, or toward
$\pm\sigma^b(\theta)$, depending on the sign of the rotation angle
$\phi$. In the OCM limit $\theta=90^\circ$ is approached, the
magnetization angle approaches $\phi=\pi/4$. In this limit there are
four degenerate Ising-type ferromagnetic states, with magnetization
either along
$\pm\sigma^a(90^\circ)$ (and $\langle\sigma^b(90^\circ)\rangle=0$), or
$\pm\sigma^b(90^\circ)$ (and $\langle\sigma^a(90^\circ)\rangle=0$).

Qualitatively the same results were obtained from the embedded
$L\times L$ clusters and they are also shown in Fig.
\ref{fig:meraMF} for comparison. While $2\times 2$ cluster is too
small and the quantum fluctuations are severely underestimated, the
two larger $3\times 3$ and $4\times 4$ clusters are qualitatively
similar and estimate the QPT point from above, see Fig.
\ref{fig:meraMF}. Rather slow convergence of these results toward
the MERA result for $\Delta E$ and for $|\phi|$ demonstrates the
importance of longer-range correlations for the correct description of
the QPT at $\theta=\theta_c$.

Altogether, these results show that the degenerate ground state of
the generalized OCM consists of a manifold of states with broken
symmetry. This confirms that the OCM is in the Ising universality
class,\cite{Mis04,Dor05} with no quantum coupling between different
broken symmetry Ising-type states. However, we found the large value
of $\theta_c\approx 84.8^{\circ}$ rather surprising and we
investigated it further using spin-wave theory. These calculations
are presented in the next Section.

Another surprise is the absence of any algebraically decaying
spin-spin correlations in the MERA ground state at $\theta_c$. They
could arise from the subleading eigenvalues
$\lambda_2,\lambda_3,\dots$ which we found to be non-zero. However,
their corresponding coefficients $c_{\alpha\beta}$ with $\alpha>1$
or $\beta>1$ in Eq. (\ref{corr}) are small (at most $\simeq10^{-4}$)
and they decay with increasing dimension $\chi$ and especially the
number of non-universal layers $N$. As a result, the only
nonvanishing term in Eq. (\ref{corr}) is the leading one for
$\alpha=\beta=1$, describing the non-decaying long range order.
Notice that this observation does not exclude non-trivial short
range correlations up to a distance $3^N$ described by the $N$
non-universal layers. We believe that when $N$ is too small, then
the missing short range correlations find a way to show up in the
small but non-zero universal coefficients $c_{\alpha\beta}$, but
these coefficients decay quickly with increasing $N$ as the short
range correlations become accurately described by the increasing
number of non-universal layers.

\section{Spin wave expansion}
\label{sec:hp}
\label{sec:lsw}

Since the spin wave expansion in powers of $1/S$ becomes exact when
the spin $S\to\infty$, we introduced a large-$S$ extension of
the generalized OCM Hamiltonian Eq. (\ref{H}) with rescaled spin
operators: $\sigma^x\rightarrow S^x/S$ and $\sigma^z\rightarrow
S^z/S$. We consider first the classical energy per site:
\begin{equation}
E_0(\theta,\phi) \equiv \langle{\cal H}(\theta)\rangle_{\phi} =
-\frac{1}{2}\left[ 1 + \cos\theta \cos(2\phi) \right]\,,
\label{Ecl}
\end{equation}
obtained using the mean-field (MF) for the ordered state of
classical spins $\vec S$, with the magnetization direction given by
Eq. (\ref{arctan}). The classical energy has a minimum at $\phi=0$
for the entire range of $\theta\in [0^\circ,90^\circ)$. However,
when the angle $\theta$ approaches $90^\circ$, the minimum becomes
more and more shallow, and finally disappears completely at
$\theta=90^\circ$. Thus, the classical ground state becomes very
sensitive to quantum fluctuations in the vicinity of the maximally
frustrated interactions in the OCM.

\begin{figure}[t!]
\includegraphics[width=7.2cm,clip=true]{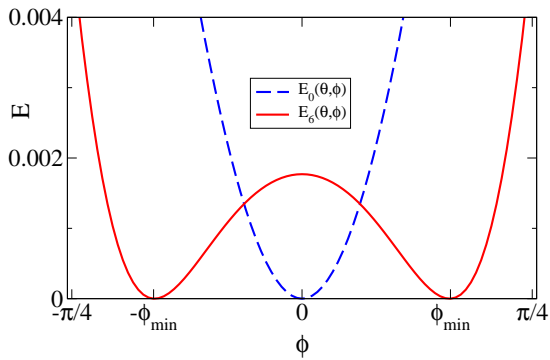}
\caption{(Color online) Mechanism of the QPT in the generalized OCM
Eq. (1) for $S=1/2$ and $\theta = 87^\circ > \theta_c$. The minimum
of the classical energy $E_0(87^\circ,\phi)$ Eq. (\ref{Ecl}) (dashed
line) at $\phi=0$ is shallow and thus unstable against weak quantum
fluctuations which induce two symmetric minima at a finite value of
$\pm\phi_{\rm min}$ obtained from $E_6(87^\circ,\phi)$ derived from
Eq.  (\ref{Hexpand}). For better comparison, $E_0$ and $E_6$ are
shifted to have a minimum \mbox{value of $0$.} } \label{fig:clqm}
\end{figure}

This behavior of the classical ground state energy explains why
{\it small\/} energy contributions due to quantum fluctuations may play
so crucial role in the generalized OCM {\it only\/} in the regime
of $\theta$ close to $90^\circ$, where they trigger a QPT by splitting
the shallow symmetric classical energy minimum at $\phi=0$ into two
symmetry-broken minima at finite values $\pm\phi_{\rm min}$ --- we
show an example of this behavior in Fig. \ref{fig:clqm} for a
particular value of $\theta>\theta_c$. Since the quantum fluctuations
induce here symmetry breaking instead of making the ground
state more symmetric, this mechanism goes beyond the Landau
functional paradigm.

We analyzed the effects of quantum fluctuations and the arising 
symmetry breaking using the Holstein-Primakoff representation of spin
$\{S_{ij}^{\alpha}\}$ operators via $\{b_{ij}\}$ bosons:
\begin{eqnarray}
\label{hp1}
\!\!\cos\phi ~ S^x_{ij} + \sin\phi ~ S^z_{ij} \!\!&=&\!
 S-b_{ij}^\dagger b_{ij}^{}\,, \\
\label{hp2}
\!\!-\sin\phi ~ S^x_{ij} + \cos\phi ~ S^z_{ij} \!\!&=&\! \frac{b_{ij}^\dag}{2}
 \sqrt{2S-b_{ij}^\dagger b_{ij}^{}} + \mathrm{H.c.}\,.
\end{eqnarray}
Operators $\{b_{ij}^{},b_{ij}^\dagger\}$ satisfy standard bosonic
commutation relations: $[b_{ij},b_{i'j'}]=0$ and
$[b_{ij}^{},b_{i'j'}^\dagger] = \delta_{ii'}\delta_{jj'}$. In this
approach, we are looking for a critical value $\theta_c$, above
which it is energetically favorable to change the direction of
magnetization $\bf{M}$ from the symmetric state $\phi=0$ to a
symmetry-broken state with a finite value of $\phi\neq 0$. We 
expanded the square root in Eq. (\ref{hp2}) in powers of $1/(2S)$ 
and obtained an expansion of Hamiltonian Eq. (\ref{H}) in powers 
of the operators $\{b_{ij}^{},b_{ij}^\dagger\}$. As we applied Wick's 
theorem to reduce the obtained Hamiltonian to an effective quadratic 
Hamiltonian, the terms proportional to the odd powers of $1/(2S)$ do 
not contribute and are skipped below (for more details see the 
Appendix). When truncated at the sixth order term this expansion reads
\begin{equation}
\widetilde{H}_6 \simeq H_0 + (2S)^{-1}H_2 + (2S)^{-2}H_4
+ (2S)^{-3}H_6\,. \label{Hexpand}
\end{equation}
Here $H_{2n}$ is a sum of all terms of the $2n$-th order in
$\{b_{ij}^{},b_{ij}^\dagger\}$ operators. In a similar way,
$\widetilde{H}_4$ and $\widetilde{H}_2$ denote expansions truncated
at the fourth and second order terms, respectively. We have found
{\it a posteriori\/} that the second order expansion
$\widetilde{H}_2$ (noninteracting spin waves) does not suffice and
higher order terms are necessary. Consequently, we consider below Hamiltonian Eq. (\ref{H}) expanded up to the sixth order.

For given $\theta$ and $\phi$, we can approximate the ground state
of the boson Hamiltonian given by Eq. (\ref{Hexpand}) by a
Bogoliubov vacuum obtained as the ground state of the quadratic
Hamiltonian $\widetilde{H}^{\rm MF}_2$ obtained using the MF
averaging of four- and six-boson terms. Details of this calculation
can be found in the Appendix \ref{sec:app}.

\begin{figure}[t!]
\includegraphics[width=7.9cm,clip=true]{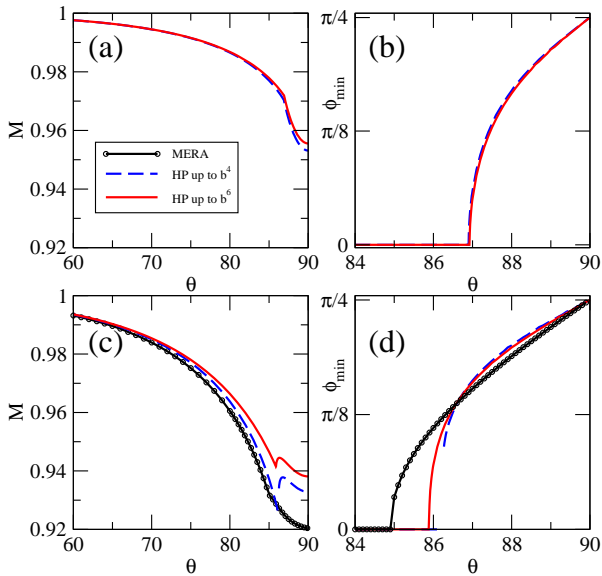}
\caption{(Color online) Symmetry breaking in the ground state as
obtained from the boson expansion Eq. (\ref{Hexpand}). Panels (a) and 
(b) show results for $S=1$, (c) and (d) --- for $S=1/2$; (a) and (c)
depict magnetization $M$ Eq. (\ref{Mlocal}), (b) and (d) --- the value 
of the magnetization angle $\phi$ Eq. (\ref{arctan}) that minimizes 
energy. Calculations for $\widetilde{H}_6$ predict the following values 
of $\theta_c$: $85.89^\circ$, $86.9^\circ$, $88.2^\circ$, and $89.2^\circ$ 
for $S=1/2$, $S=1$, $S=2$, and $S=5$, respectively (the last two not
shown), and $\theta_c\to 90^\circ$ for $S\to\infty$. }
\label{fig:HP}
\end{figure}

First we performed separate calculations for $\widetilde{H}_2$,
$\widetilde{H}_4$ and $\widetilde{H}_6$ for several values of spin
$S\geq 1$ when the $1/(2S)$-expansion given in Eq. (\ref{Hexpand})
is convergent. The quadratic $\widetilde{H}_2$ fails for large
$\theta$, where its spectrum becomes gapless and the magnetization
$M$ Eq. (\ref{Mlocal}) diverges. In contrast, $\widetilde{H}_4$ and
$\widetilde{H}_6$ give only small reduction in $M$ in the entire
range of $\theta$, see Fig. \ref{fig:HP}(a) and \ref{fig:HP}(c).
Interestingly, the Bogoliubov spectrum remains gapful at $\theta_c$
in both the fourth and sixth order expansions and, just like in the
MERA, there are no algebraically decaying spin-spin correlations.
The critical angle $\theta_c$ at which the symmetry-breaking QPT
occurs increases toward $90^\circ$ with increasing $S$ when the
quantum fluctuations become less significant. Therefore, the
magnetization $M$ increases with increasing $S$ and it tends to $1$
in the classical limit $S\to\infty$.

Encouraged by these results, we also performed similar calculations
for the generalized OCM Eq. (\ref{H}) with $S=1/2$, where the
convergence of the $1/(2S)$-expansion becomes problematic. Unlike
for $S\geq1$, we find that the fourth order expansion is
insufficient as it predicts the first order QPT [Fig.
\ref{fig:HP}(d)] and does not agree qualitatively with the
prediction of the MERA, see Sec. \ref{sec:num}. Only in the sixth
order one finds a qualitative agreement between the present boson
expansion and the MERA, both giving the second order QPT at
$\theta_c$. A cusp in $M(\theta)$ seen in Fig. \ref{fig:HP}(c) shows
that even the sixth order expansion is not quite converged for
$S=1/2$. Again, the Bogoliubov spectrum remains gapful at $\theta_c$
in the sixth order expansion, with a finite gap equal $1.52$, and
one finds no algebraically decaying spin-spin correlations.

\section{Conclusions}
\label{sec:summa}

Summarizing, we found that a second order quantum phase transition
in the generalized orbital compass model Eq. (\ref{H}) occurs at
$\theta_c=84.8^\circ$ which is surprisingly close to the compass
point $\theta=90^\circ$, i.e., only when the interactions are
sufficiently strongly frustrated. There is spontaneous ferromagnetic
magnetization at any angle $\theta\in[0^\circ,90^\circ]$. Below
$\theta_c$ the ferromagnetic ground state is doubly degenerate with
the spontaneous magnetization, either parallel or antiparallel to
the average direction $\sigma^a_{ij}+\sigma^b_{ij}$. None of the
directions, neither $a$ nor $b$, is preferred in this symmetric
phase. In contrast, when $\theta$ increases above $\theta_c$ the
symmetry between $a$ and $b$ becomes spontaneously broken and the
ferromagnetic magnetization begins to align parallel/antiparallel to
either $\sigma^a_{ij}$ or $\sigma^b_{ij}$. The ground state is
fourfold degenerate in this symmetry-broken phase. The spontaneous
magnetization $M$ is close to $1$ and quantum fluctuations remain
small in the whole range of $\theta\in(0^\circ,90^\circ]$.

These results were obtained using the MERA and the mechanism of the
QPT was explained within the spin-wave theory. For classical spins
the minimum of energy is at one of the two symmetric states with the
magnetization either parallel or antiparallel to
$\sigma^a_{ij}+\sigma^b_{ij}$, see Fig. \ref{fig:clqm}. The minimum 
becomes more and more shallow as the compass point $\theta=90^\circ$ 
is approached. However, the quantum fluctuations are weak due to the 
gapful orbital wave excitations, and only very close to the above 
OCM point become strong enough to split the shallow minimum into two 
distinct minima in the vicinity of the OCM point. In this way the 
symmetry between the axes $a$ and $b$ is spontaneously broken. For 
this reason the orbital $e_g$ model with ferro-orbital interactions, 
considered in Ref. \onlinecite{vdB99} and corresponding to a 
``moderate'' value of $\theta=60^\circ$ [see Fig. \ref{fig:ocm}(b)], 
orders in a symmetric (uniform) phase induced by the stronger 
(here $\propto\sigma^x_{ij}\sigma^x_{i'j'}$) interaction component.

Interestingly, since --- unlike in the Landau paradigm --- the symmetry
in the present model Eq. (\ref{H}) is broken rather than restored by
quantum fluctuations, we do not find
any algebraically decaying spin-spin correlations at the critical point
found in the generalized orbital compass model Eq. (\ref{H}). The spin 
waves also remain gapful at this point.

\acknowledgments

We thank P. Horsch for insightful discussions. L.C., J.D. and A.M.O.
acknowledge support by Polish Ministry of Science and Higher Education 
under Projects No. N202 175935, N202 124736 and N202 069639, respectively. 
A.M.O. was also supported by the Foundation for Polish Science (FNP).

\appendix*
\section{Details of the spin wave calculation}
\label{sec:app}

In this section we present the details of the spin wave calculation.
We consider the general case of an $L\times L$ square lattice, with
$L$ being odd for convenience. The results presented in Section
\ref{sec:hp} are obtained after taking the thermodynamic limit
$L\to\infty$.

For given $\theta$ and $\phi$, the ground state of the boson
Hamiltonian Eq. (\ref{Hexpand}) is approximated by a Bogoliubov
vacuum obtained as the ground state of a mean-field (MF) quadratic
Hamiltonian $\widetilde{H}^{\rm MF}_2$ (to be derived later on).
Terms $H_2$, $H_4$ and $H_6$ in Eq. (\ref{Hexpand}) are given by:
\begin{multline}
H_2 ~=~
4[1+\cos \theta \cos (2\phi)] \sum_{\bf r} b^\dagger_{\bf r}b_{\bf r} \\
- \sin^2 \left( \phi - \frac{\theta}{2} \right)
\sum_{\bf r}\left( b^\dagger_{\bf r} b_{{\bf r} + {\bf e_x}} + b_{\bf r}b_{{\bf r}
+ {\bf e_x}} + \mathrm{H.c.}\right) \\
- \sin^2 \left( \phi + \frac{\theta}{2} \right)
\sum_{\bf r}\left( b^\dagger_{\bf r} b_{{\bf r} + {\bf e_y}} + b_{\bf r}b_{{\bf r}
+ {\bf e_y}} + \mathrm{H.c.} \right) \, ,
\end{multline}
\begin{multline}
H_4 ~=~
-4 \cos^2 \left( \phi - \frac{\theta}{2} \right) \sum_{\bf r}
b^\dagger_{\bf r} b^\dagger_{{\bf r} + {\bf e_x}} b_{\bf r} b_{{\bf r}
+ {\bf e_x}} \\
-4 \cos^2 \left( \phi + \frac{\theta}{2} \right) \sum_{\bf r}
b^\dagger_{\bf r} b^\dagger_{{\bf r} + {\bf e_y}} b_{\bf r} b_{{\bf r}
+ {\bf e_y}} \\
+ \frac{1}{2} \sin^2 \left( \phi - \frac{\theta}{2} \right) \sum_{\bf r}
\left\{ b^\dagger_{\bf r} b_{\bf r}^2 (b_{{\bf r} \pm {\bf e_x}}
+ b^\dagger_{{\bf r} \pm {\bf e_x}}) + \mathrm{H.c.} \right\} \\
+ \frac{1}{2} \sin^2 \left( \phi + \frac{\theta}{2} \right) \sum_{\bf r}
\left\{ b^\dagger_{\bf r} b_{\bf r}^2 (b_{{\bf r} \pm {\bf e_y}}
+ b^\dagger_{{\bf r} \pm {\bf e_y}}) + \mathrm{H.c.} \right\} \, ,
 \label{H4}
\end{multline}
\begin{multline}
H_6 ~=~
\frac{1}{8} \sin^2 \left( \phi - \frac{\theta}{2} \right)
 \sum_{\bf r} \left\{  \left( b^\dagger_{\bf r} b_{\bf r} \right)^2 b_{\bf r} ( b_{{\bf r} \pm  {\bf e_x}} + b^\dagger_{{\bf r} \pm  {\bf e_x}}) \right. \\
\left.  - ~ 2 b^\dagger_{\bf r} b^\dagger_{{\bf r} +  {\bf e_x}} b_{\bf r}^2 ( b_{{\bf r} +  {\bf e_x}} + b^\dagger_{{\bf r} +  {\bf e_x}}) b_{{\bf r} +  {\bf e_x}} + \mathrm{H.c.} \right\} \\
+ \frac{1}{8} \sin^2 \left( \phi + \frac{\theta}{2} \right)
 \sum_{\bf r} \left\{  \left( b^\dagger_{\bf r} b_{\bf r} \right)^2 b_{\bf r} ( b_{{\bf r} \pm  {\bf e_y}} + b^\dagger_{{\bf r} \pm  {\bf e_y}}) \right. \\
\left.  - ~ 2 b^\dagger_{\bf r} b^\dagger_{{\bf r} +  {\bf e_y}} b_{\bf r}^2 ( b_{{\bf r} +  {\bf e_y}} + b^\dagger_{{\bf r} +  {\bf e_y}}) b_{{\bf r} +  {\bf e_y}} + \mathrm{H.c.} \right\} \, ,
\end{multline}
where  ${\bf r} = (i,j)$, ${\bf e_x} = (1,0)$ and ${\bf e_y} = (0,1)$. The $\pm$ signs mean here that both terms, with $+$ and $-$ sign separately, must be taken into account.

To derive the quadratic approximation $\widetilde{H}^{\rm MF}_2$, we
replace the boson terms in $H_4$ and $H_6$ with two-boson terms and
proper averages by means of the MF approximation and Wick's theorem.
This justifies {\it a posteriori\/} why the expansion
(\ref{Hexpand}) is limited only to the terms with {\it even\/}
number of boson operators. As an example of this approximation,
consider one of the contributions to $H_4$ in Eq. (\ref{H4}):
$b_{\bf r}^\dagger b_{\bf r}^2 b_{{\bf r} + {\bf e_x}}$, which is
replaced with a quadratic term:
\begin{multline}
b_{\bf r}^\dagger b_{\bf r} ^2 b_{{\bf r} + {\bf e_x}}
~\simeq~
2 \langle b_{\bf r}^\dagger b_{\bf r} \rangle \; b_{\bf r} b_{{\bf r} + {\bf e_x}} +
2 \langle b_{\bf r} b_{{\bf r} + {\bf e_x}} \rangle \; b_{\bf r}^\dagger b_{{\bf r}} \\
+ \langle b_{\bf r}^\dagger b_{{\bf r} + {\bf e_x}} \rangle \; b_{\bf r} ^2 +
\langle b_{\bf r}^2 \rangle \; b_{\bf r}^\dagger b_{{\bf r} + {\bf e_x}} -
\langle b_{\bf r}^\dagger b_{\bf r} ^2 b_{{\bf r} + {\bf e_x}} \rangle \, .
\end{multline}

The above replacement procedure leads to six MF parameters $\{m_i\}
\equiv\{m_1,m_2,\ldots,m_6\}$ that should satisfy self-consistency
conditions. These are in fact all possible combinations of operators
defined on nearest-neighbor sites that cannot be derived one from
another by commutation relations and translational invariance of the
lattice, i.e.,
$m_1 = \langle b^\dagger_{\bf r} b_{\bf r}\rangle$, $m_2 =
\langle b^\dagger_{\bf r} b_{{\bf r} + {\bf e_x}}\rangle$, $m_3 = \langle
b^\dagger_{\bf r} b_{{\bf r} + {\bf e_y}}\rangle$, $m_4 = \langle b_{\bf
r}^2\rangle$, $m_5 = \langle b_{\bf r} b_{{\bf r} + {\bf
e_x}}\rangle$, and $m_6 = \langle b_{\bf r} b_{{\bf r} + {\bf
e_y}}\rangle$.

The obtained Hamiltonian $\widetilde{H}^{\rm MF}_2$ is diagonalized
by the Fourier transformation followed by the Bogoliubov transformation.
Fourier transformation which is consistent with periodic boundary
conditions $b_{L+1,j} = b_{1,j}$ and $b_{i,L+1} = b_{i,1}$ has the
following form:
\begin{equation}
b_{\bf r} ~=~ \frac{1}{L} \sum_{\bf k} b_{\bf k} \: e^{i \, {\bf k} \cdot {\bf r}}\,,
\end{equation}
where ${\bf k} = (k_x,k_y)$ is the momentum. In the sum, momentum
components $k_x$ and $k_y$ take the values (for odd $L$ considered here):
\begin{equation}
k_{x(y)} ~=~ 0 \cdot \frac{2\pi}{L}, \; \pm 1 \cdot \frac{2\pi}{L},
\; \ldots ,\; \pm \frac{L-1}{2} \cdot \frac{2\pi}{L} \,.
\end{equation}
Diagonalization of $\widetilde{H}^{\rm MF}_2$ is completed by the
Bogoliubov transformation:
\begin{equation}
b_{\bf k} ~=~ u_{\bf k}\gamma_{\bf k} + v_{-\bf k}^\ast \gamma_{-\bf k}^\dagger \, ,
\end{equation}
where the modes $u_{\bf k}$ and $v_{\bf k}$ are normalized such that
$|u_{\bf k}|^2 - |v_{\bf k}|^2 = 1$. The obtained modes are used to
calculate new values of the MF parameters $\{m_i\}$
($i=1,2,\ldots,6$). For instance, one of them reads: $m_2=\langle
b^\dagger_{\bf r} b^{}_{{\bf r} + {\bf e_x}}\rangle =
\frac{1}{L^2}\sum_{\bf k} |v_{\bf k}|^2 \cos k_x$. Starting from
random values, the above steps are iteratively applied until full
convergence of all $\{m_i\}$ is reached, which results in satisfying
the self-consistency conditions.



\begin{thebibliography}{99}
\bibitem{notefo} The orbital interactions in transition metal oxides
                 favor alternating orbital order, see Refs. 
                 \onlinecite{Tok00,Hor08} --- the corresponding OCM is 
                 equivalent to the present ferro-orbital model Eq. 
                 (\ref{H}) by a unitary transformation.
\bibitem{note} The comparison has been made for geometries with the same bond dimension in all layers. The performance of 5-to-1 geometry can be further improved by considering different bond dimensions on some indices.
\end{thebibliography}
\end{document}